\documentclass[aps,superscriptaddress,nofootinbib,eqsecnum,pra,notitlepage,twocolumn]{revtex4-1} 

\usepackage{amsfonts}
\usepackage{amsmath}
\usepackage{amssymb}
\usepackage{graphicx,color}
\usepackage{float}
\usepackage{hyperref}
\usepackage{subfigure}
\usepackage{dcolumn}
\usepackage{soul}
\usepackage{ulem}
\usepackage{verbatim}

\DeclareMathAlphabet{\mathpzc}{OT1}{pzc}{m}{it}

\begin{document}

 \title{Two heavy impurities immersed in  light few-boson systems for non-integer dimensions}

\author{R. M. Francisco}

\affiliation{Universidade Federal de S\~{a}o Jo\~{a}o del Rei, DCNAT, 36301-160 S\~{a}o Jo\~{a}o del Rei, MG, Brazil}

 \author{D. S. Rosa}

\affiliation{Instituto Tecnol\'{o}gico de Aeron\'{a}utica, DCTA,
  12228-900 S\~{a}o Jos\'{e} dos Campos, SP, Brazil}
 
 \author{T. Frederico}

\affiliation{Instituto Tecnol\'{o}gico de Aeron\'{a}utica, DCTA,
  12228-900 S\~{a}o Jos\'{e} dos Campos, SP, Brazil}

\begin{abstract}
 We investigate the spectrum and structure of two heavy bosonic impurities immersed in a light boson system in non-integer dimensions  $(D)$  by means of the Born-Oppenheimer approximation. The ratio between the energies of two consecutive many-body bound states, which follows an Efimov type geometrical scaling law when the heavy-light pair has a resonant s-wave interaction, is found to be a function of the mass asymmetry, number of light bosons and effective dimension $D$. Furthermore, the wave function is computed and its behavior explored for different system configurations. To exemplify our results, we consider mixtures of two heavy caesium atoms interacting with up to two lithium ones, which are systems of current experimental interest.  
\end{abstract}

\maketitle

\section{INTRODUCTION}

The Efimov effect consists in the appearance
of an infinite tower of three-body bound states that follows an universal geometrical scaling law for resonant short-range two-body forces~\cite{efimov}. This effect was originally proposed in the nuclear physics context, however, its first evidence was found by Kraemer et al.~\cite{kraemer} in trapped ultracold atomic systems. In their work, the Feshbach resonance technique was applied to an ultracold gas of caesium atoms, where they observed three-body recombination losses when the strength of the two-body interaction was varied and the weakly bound Efimov state turned into a continuum resonance (see also~\cite{homoexp1,homoexp2,heteexp1,heteexp2,heteexp3}). Nowadays, it is known that the Efimov effect is present in several systems, such as atomic gases~\cite{efimovatoms}, polarons~\cite{efimovpolarons}, dipolar molecules~\cite{efimovdipolar}, strong interacting photons~\cite{efimovphotons} and in general atomic and nuclear physics contexts \cite{nielsen,Braaten:2004rn,Frederico:2012xh,Greene:2017cik,Naidon:2016dpf}.

The Efimov effect is closely related to the collapse of the three-body bound energy discovered by Thomas in 1935~\cite{thomas}. By decreasing the range of the interaction, Thomas found that the three-body bound state problem admits a solution at any energy with a spectrum “unbounded from below”. Both phenomena are affected by changes in the effective dimension $D$ in which the system is embedded. For example, just as the Thomas collapse, the Efimov effect is absent in $D = 2$. It has been theoretically demonstrated for homonuclear systems that the Efimov effect survives only for dimensions in the range $2.3 < D < 3.8$~\cite{nielsen}. Considering heteronuclear systems, recent works have shown how the mass imbalance changes the range of dimensions where the Efimov effect is present~\cite{rosastm,mohapatra,cristensen}. 

Although technically simple to implement, the $D$-dimensional approach presents a key issue, i.e., the connection of the dimension $D$ to the geometry properties of traps used in experiments. Recently, it was suggested that a relation between
non-integer dimensions and the trap deformation induced by an external
harmonic potential could be given by
\begin{equation}
3(D-2)/(3-D)(D-1)= b_{ho}^2/r_{2D}^2\,,
\label{eq:Dtrap}
\end{equation} 
where  $b_{ho}$ is the oscillator
length and $r_{2D}$ is the root mean square radius of the three-body system in two dimensions~\cite{garridoprr}. 
This relation was found by associating the three-body bound state energies obtained within two different contexts, that is, by squeezing the system in an external harmonic potential and by embedding it in $D$ dimensions. The validity of the connection was tested for Gaussian and Morse short-range two-body potentials showing the universal nature of the connection regardless to the details of the two-body potential. It is worth mentioning that this translation  has not yet been computed for a larger number of
particles. Despite the advances that led to the possibility to compress and expand the atomic cloud, creating effectively two-~\cite{BEC2D}  and one-dimensional~\cite{BEC1D} setups, to the best of our knowledge, there are not yet experiments
designed to measure the effect of continuously changing the trap geometry in the geometrical scaling parameter.

Besides the Efimov effect, in literature we find studies dedicated to the universal properties of systems with more than three particles at the unitary limit, where the dimer binding energy vanishes, or, equivalently, the scattering 
length is driven to infinity ~\cite{kroger,adhikari,naus,yamashita81,yamashita75,hadizadeh2011,von,yan,NaidonFBS2018}. In these studies, universal correlations between the binding energies of successive tetramers was discovered by solving the Faddeev-Yakubovsky (FY) equations. The findings corroborate the existence of a four-body limit-cycle beyond the Efimov one~\cite{hadizadeh2011}. Following that, the shift in the four-body recombination peaks due to the finite length corrections was studied~\cite{hadizadeh2013}, explaining experimental observations for ultracold gases of caesium atoms. Recently, the independence between few-body scales was demonstrated for many-boson systems composed by two heavy atoms interacting with ($N-2$) light ones at the unitary limit~\cite{interwoven}, showing different limit-cycles, each one associated with a given number of bosons present in the system. 

In the present work, we extend the investigation of the many-boson system in reference~\cite{interwoven} by embedding it in an environment defined by an effective dimension $D$. The mass imbalance and the
number of light atoms are also varied to study how these changes
impact the geometrical ratio of consecutive many-body bound states. In addition to results presented at the unitary limit, we also analyse finite-range corrections for the many-body spectrum and structure. To accomplish this, we obtain an expression for the two-body bound energy as a function of the scattering length and effective dimension $D$, which can also provide a more directly way to link theoretical to experimental results. Besides that, we obtain the wave function of the two heavy atoms, as well as an analytical expression for the wave function of the ($N-2$) lighter ones, where $N$ is the total number of atoms in the system. In the case where it applies, we associate the effective dimension $D$ with the trap deformation.

The work is organized as follows. In section~\ref{sec:BOA}, we present the Born-Oppenheimer formalism used to solve the bound state problem of two heavy impurities within the system of light bosons embedded in an effective dimension $D$. Besides that, we analytically compute the $(N-2)$ light bosons wave function, which can be used to parameterize the many-body large-distance tail of the total wave function. Furthermore, we explicitly write the scaling parameter related with successive ratios of many-body bound states for different mass ratios and dimensions considering zero bound heavy-light energies. For three-body systems, the results are compared with the ones found by solving the Skorniakov and Ter-Martirosyan equations. In section~\ref{ImpBoundStates}, for the two heavy impurities, we present bound state energies and wave functions considering corrections of finite heavy-light bound energies. We focus in the limit of discrete scale symmetry to the point where the continuum one takes place. Correlations between the many-body bound state energies as functions of dimension and mass imbalance are discussed as possible limit-cycles. Finally, we apply the present approach to molecules of $^6$Li$_{N-2}-\,^{133}$Cs$_2$ for $N=3$ and $N=4$. In section~\ref{conclusion}, we summarize our work.

\section{BORN-OPPENHEIMER approach}\label{sec:BOA}

\begin{center}
\begin{figure}[!htb]
\includegraphics[width=8.5cm]{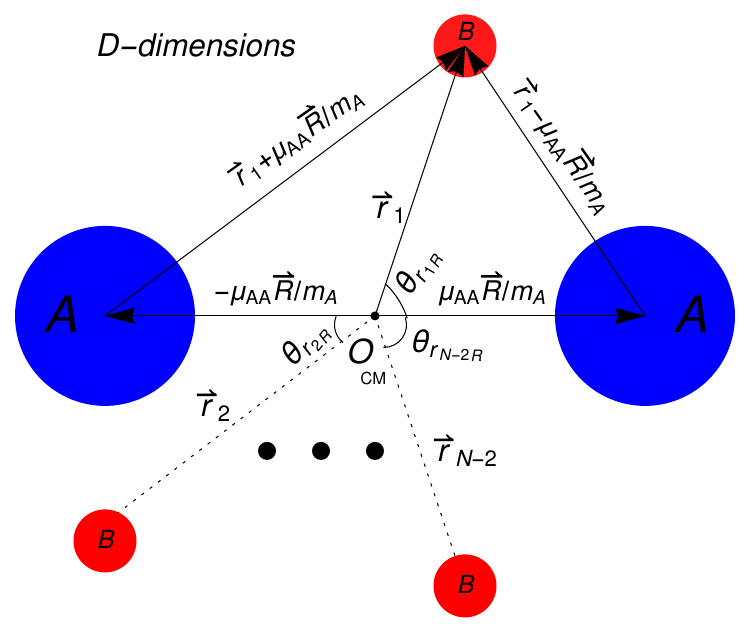}
\caption{ Representation of the AABB...B structure forming the $N$-body system composed by two identical heavy-atoms with
masses $m_A$, and ($N-2$) light-atoms with masses $m_B$. }
\label{fig1}
\end{figure}
\end{center}

In this section, we present the formalism used to study the many-body system embedded in $D$-dimensions composed by two heavy particles with masses $m_A$ and ($N-2$) light bosonic atoms with masses $m_B$. This system is depicted in figure~\ref{fig1}.

%%%%%%%%%%%%%%%%FIGURE01%%%%%%%%%%%%%%%%%%%
%%%%%%%%%%%%%%%%%%%%%%%%%%%%%%%%%%%%%%%%%%%%%

We are interested in the relative motion between the particles of the system, so that, ignoring the movement of the center of mass, the Hamiltonian can be written in relative coordinates as
\begin{eqnarray} \label{eq:hamiltonian}
&&H= -\frac{\hbar^{2}}{2\mu_{AA}}\nabla^{2}_{R} +V_{B}(|\textbf{R}|)+ \sum_{i=1}^{N-2}\left(-\frac{\hbar^{2}}{2\mu_{B,AA}}\nabla^{2}_{r_i}  \right. \nonumber \\
&&+\left. \sum_{j=1}^{2}V_{A}\left(\Big\vert\textbf{r}_i+(-1)^{j}\frac{\mu_{AA}}{m_A}\textbf{R}\Big\vert\right)\right),
\end{eqnarray}
where $\mu_{AA} = m_{A}/2$ and $\mu_{B,AA} = 2m_A m_B /(2m_A +m_B )$ are the reduced masses of the system, while $V_A$ and $V_B$ denote the $AB$ and $AA$ two-body interactions, respectively. 

In order to apply the BO approximation, let us consider $m_B \ll m_A$. Within this approach, the total wave function of the system  is given by a product between the wave functions of the $N-2$  light fast atoms ($\Phi$) and of the heavy slow ones ($\phi$), which can be written as
\begin{eqnarray}
\Psi(\textbf{r}_1,\textbf{r}_2,...,\textbf{r}_{N-2},\textbf{R}) = \phi(\textbf{R}) \Phi_R(\textbf{r}_1,\textbf{r}_2,...,\textbf{r}_{N-2})\, .\ \ \ \ \ \
\end{eqnarray}
Since there is no interaction between the light bosons, we assume that the wave function of the light atoms is a product of single particle states, given by
\begin{equation}\label{eq:productwf}
 \Phi_R(\textbf{r}_1,\textbf{r}_2,...,\textbf{r}_{N-2})\,=\prod_{i=1}^{N-2}\psi_R(\textbf{r}_{i})\, .
\end{equation}

In the lowest order of approximation, the action of the Laplace operator $\nabla_R^{2}$ on the total wave function $\Psi$ can be written as 
\begin{equation}
\nabla_R^{2}\left[\phi(\textbf{R})\prod_{i=1}^{N-2}\psi_R(\textbf{r}_{i}) \right]\approx \prod_{i=1}^{N-2}\psi_R(\textbf{r}_{i})  \nabla_R^{2}\phi(\textbf{R})\, .
\end{equation}
Here, it is possible to make this approximation because the other terms generated by the operator action are suppressed by the mass factor ratio. By adopting the BO approximation and our particular choice of Hamiltonian where the light-light potential vanishes, we can write two independent equations for each light particle and for the two heavy ones as
\begin{multline}
\left[-\frac{\hbar^{2}}{2\mu_{B,AA}}\nabla^{2}_{r_i}  + \sum_{j=1}^{2}V_{A}\left(\Big\vert\textbf{r}_i+(-1)^{j}\frac{\textbf{R}}{2}\Big\vert\right)\right]\psi_R(\textbf{r}_{i})  \\
=\epsilon(R)\psi_R(\textbf{r}_{i})\,, 
\label{eqlightatom}
\end{multline}
and
\begin{multline}
\left[- \frac{\hbar^{2}}{2\mu_{AA}} \nabla_{R}^{2}+ V_{B}(|\textbf{R}|)+(N-2)\epsilon(R) \right] \phi(\textbf{R}) \\
 = E_{N} \phi(\textbf{R})\,,\label{eqheavy}
\end{multline}
respectively. The eigenvalue $\epsilon(R)$ of the Eq.~\eqref{eqlightatom} depends on the
relative position of the heavy atoms and enters as an
effective potential in Eq.~\eqref{eqheavy}, while the eigenvalue $E_{N}$ of the heavy atoms equation gives the many-body energy.

As the total wave function is symmetric under the interchange of the 
heavy atoms and the interaction is not dependent on the spin, the formalism developed here is suitable to describe many-body systems formed by bosonic or fermionic heavy impurities, being the last one in antisymmetric spin states.

\section{Light atoms structure}

In the present model, as we have described, the zero-range potential acts only in the heavy-light pairs, which allows us to use the analytical expression for $\epsilon(R)$ obtained in~\cite{rosaBO}, where three-body systems in $D$-dimensions were studied. Within this approach, the energy eigenvalue of the light bosons comes from the solution of a transcendental equation, given by
\begin{multline}
2^\frac D2\left(\frac{\sqrt{|\overline{\epsilon}({\overline R})|}}{{\overline R}}\right)^\frac{D-2}{2}
K_\frac{D-2}{2}\left({\overline R}\sqrt{|\overline{\epsilon}({\overline R})|}\right)\\
-  
\frac{\pi\csc\left(D\pi/2\right) }{\Gamma(D/2)} \,
\left(1-|\overline{\epsilon}
({\overline R})|^\frac{D-2}{2}\right) = 0\,,
\label{fullbopot}
\end{multline}
where $|\overline{\epsilon}(\overline{R})| = |\epsilon(\overline{R})|/|E^{(D)}_{2}|$  and $|E^{(D)}_{2}|$ is the two-body bound state energy of each pair of heavy-light atoms. The dimensionless relative distance between the heavy atoms is ${\overline R}=R/L$, with $L=\sqrt{\hbar^{2}/{2\mu_{B,AA}|E^{(D)}_{2}|}}$.  $K_\alpha(z)$ and $\Gamma(z)$ are the modified Bessel function of the second kind and the gamma function, respectively.

%%%%%%%%%%%%%%%%FIGURE02%%%%%%%%%%%%%%%%%%%
\begin{center}
\begin{figure}[!thb]
\includegraphics[width=8.5cm]{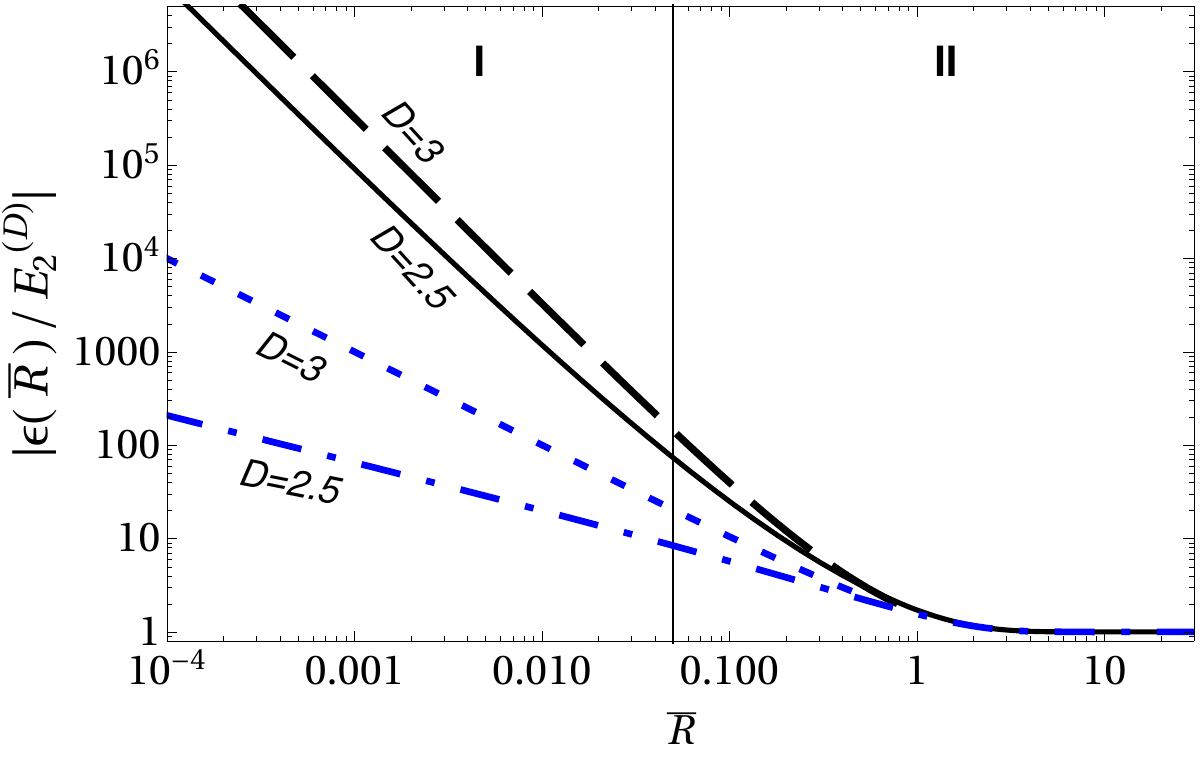}
\caption{ Effective potential between the two heavy atoms
as a function of the dimensionless  distance,  $\overline R=R/L$ $(L=[ \hbar^{2}/{2\mu_{B,AA}|E^{(D)}_{2}|}]^\frac12)$. The results for the effective potential from the solution of Eq.~\eqref{fullbopot} are presented for $D=3$ (black long-dashed line)
and $D=2.5$ (black solid line).
The  asymptotic form of the potential at large distances $(\overline R>>1)$ from Eq.~\eqref{assymplarge} are shown for $D=3$ (blue short-dashed line) and  $D=2.5$ (blue dot-dashed line). Regions I and II are  separated by the vertical line and  region I corresponds to $\overline R<<1$ where the potential is dominated by the asymptotic form  given in Eq.~\eqref{assympsmall}  with 
$g(D=3)=-0.322$  and with
 $g(D=2.5)=-0.081$.  }
\label{fig2}
\end{figure}
\end{center}
%%%%%%%%%%%%%%%%%%%%%%%%%%%%%%%%%%%%%%%%%%%%%

The effective potential between the heavy particles from the solution of Eq.~\eqref{fullbopot} assumes quite simple forms in the limits of large and small distances $\overline R>>1$ and $\overline R<<1$. These forms are given by
\begin{multline}
\lim_{R\rightarrow \infty}|\epsilon(R)| = |E^{(D)}_{2}| \\+|E^{(D)}_{2}| K_\frac{D-2}{2}\left(\overline R \right)  
\frac{2^{\frac{(D+2)}{2}}\Gamma(D/2)\,\overline R \ ^{\frac{2-D}{2}}}{(2-D)\pi \csc({D\pi/2})} \, ,
\label{assymplarge}
\end{multline}
and 
\begin{equation}
\lim_{R\rightarrow 0}|\epsilon(R)| = \frac{\hbar^{2}\,g(D)}{2\mu_{B,AA}R^{2}} \, ,
\label{assympsmall}
 \end{equation}
where $g(D)$ is the solution of the transcendental equation
\begin{equation}\label{eq:gD}
g(D) = \left[-\frac{ \pi \csc(D \pi/2)}
{ 2^\frac{D}{2} \Gamma({D}/{2}) K_\frac{D-2}{2}\big(\sqrt{g(D)}\big) } \right]^\frac{4}{2-D}.
\end{equation}

At the unitary limit, where the heavy-light binding energy vanishes for any finite distance between the heavy atoms, only the contribution proportional to $1/R^2$ from Eq.~\eqref{assympsmall} survives. This form of the potential allows the presence of the Thomas collapse for dimensions $2<D<4$, restricted to conditions of mass configuration and effective dimension $D$. For illustration, in Fig.\ref{fig2} we plot the effective potential between the two heavy atoms for two different dimensions. Basically, the value of the two-body energy defines the tail of the potential and the relative distance in which the two heavy atoms are under the action of the inverse square interaction. In the unitary limit, region $II$ disappears, and we have the $g(D)/R^2$ behaviour of the effective potential represented in region $I$.

We will now turn our attention to the wave function. The Fourier transform of $\psi_R(\textbf{r}_i)$ can be analytically calculated from
\begin{equation}
 \psi_{R}(\textbf{r}_i) = \int \frac{d^{D}p_{i}}{(2\pi)^{D}}\ \frac{ e^{i \textbf{p}_{i}.(\textbf{r}_{i}+\textbf{R}/2 ) } +
 e^{i\textbf{p}_{i}.(\textbf{r}_{i} - \textbf{R}/2)} }
 {\epsilon(R)-p_{i}^{2}/2\mu_{B,AA}},
 \end{equation}
which gives
\begin{eqnarray}
&&\psi_{R}(\textbf{r}_{i}) = - \frac{2\mu_{B,AA}}{\hbar^{2}(2\pi)^\frac{D}{2}}
\left(\sqrt{ \frac{2\mu_{B,AA}|\epsilon(R)|} {\hbar^{2}}} \right)^\frac{D-2}{2} \nonumber \\
&&\times
\left[\left( \bigg\vert \textbf{r}_{i}+\frac{\textbf{R}}{2}\bigg\vert \right)^\frac{2-D}{2} K_\frac{D-2}{2}\left( \sqrt{\frac{2\mu_{B,AA}|\epsilon(R)|} {\hbar^{2}} }\bigg\vert \textbf{r}_{i}+\frac{\textbf{R}}{2} \bigg\vert \right)\right. \nonumber \\
&&+\left.\left( \bigg\vert \textbf{r}_{i}-\frac{\textbf{R}}{2}\bigg\vert \right)^\frac{2-D}{2} K_\frac{D-2}{2}\left(\sqrt{ \frac{2\mu_{B,AA}|\epsilon(R)|} {\hbar^{2}}} \bigg\vert \textbf{r}_{i}-\frac{\textbf{R}}{2} \bigg\vert  \right)\right]. \nonumber \\
\label{fullwavefunciton}
\end{eqnarray}
 In the unitary limit, where $ E^{(D)}_2$ goes to $0$ and $R\gg \hbar/\sqrt{m\,E^{(D)}_2}$, $|\epsilon(R)|$ is given by Eq.~\eqref{assympsmall} and the wave function written in Eq.~\eqref{fullwavefunciton} is given by
\begin{multline}
\psi_{R}(r_{i}) = - \frac{R^{2-D}}{(2\pi)^\frac{D}{2}}\frac{2\mu_{B,AA}}{\hbar^{2}}
g(D)^\frac{D-2}{4}  \\
\times
\left[ \zeta_{+}^\frac{2-D}{4} K_\frac{D-2}{2}\left( \sqrt{g(D) \zeta_{+}} \right)
+\zeta_{-}^\frac{2-D}{4} K_\frac{D-2}{2}\left( \sqrt{g(D) \zeta_{-}} \right)\right], \\
\end{multline}
where
\begin{eqnarray}
 \zeta_{\pm}=\frac{r_i^2}{R^2}+\frac{1}{4}\pm\frac{r_i}{R}\cos(\theta_{r_i R}).
\end{eqnarray}

%%%%%%%%%%%%%%%%FIGURE03%%%%%%%%%%%%%%%%%%%
\begin{center}
\begin{figure}[!htb]
\subfigure[]{\includegraphics[width=8.2cm]{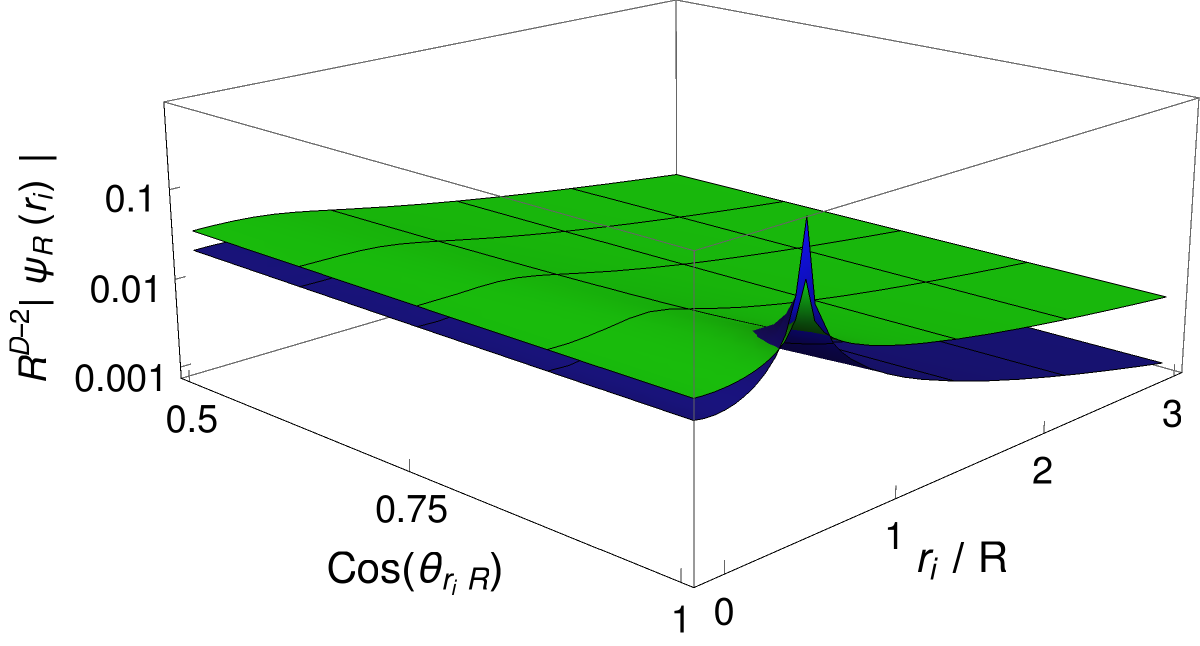}}
\subfigure[]{\includegraphics[width=8.2cm]{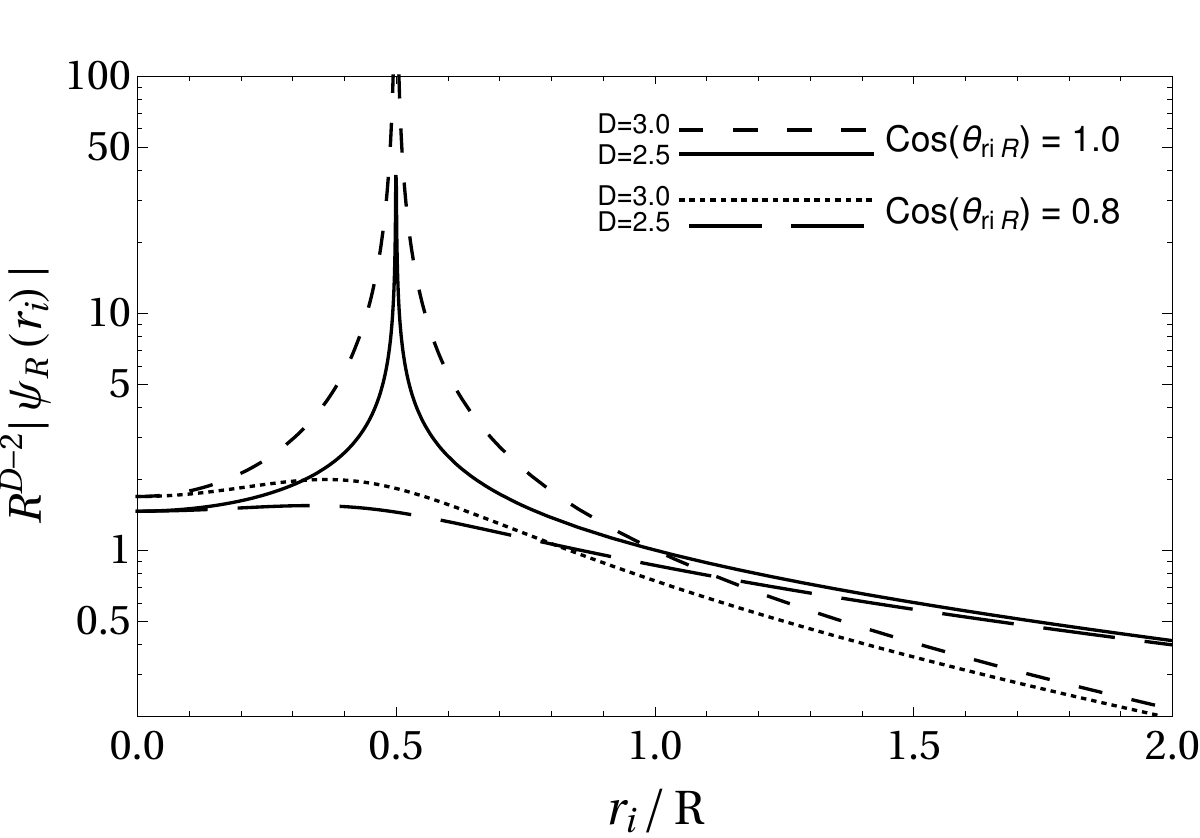}}
\caption{ In units of $\hbar=m_A=1$, we show in figure (a): the wave  function of each light-atom in the system $^6$Li$_{N-2}$-$^{133}$Cs$_2$ considering two different dimensions $D=3$ (blue surfaces) and $D=2.5$ (green surface) or, in the last case, trap geometry $ b_{ho}/r_{2D}=1.414$. Figure (b): wave function for fixed $\cos \theta_{r_iR}$ and different dimensions as a function of the ratio $r_i/R$.}
\label{fig3}
\end{figure}
\end{center}
%%%%%%%%%%%%%%%%%%%%%%%%%%%%%%%%%%%%%%%%%%%%%

 In figure~\ref{fig3}, embedded in two different dimensions, namely $D=3$ and 2.5, we show the light particle wave function at the unitary limit considering a system composed by $^6$Li and $^{133}$Cs. In  figure \ref{fig3}~(a), the peak at $r_i/R=1/2$ and $\cos(\theta_{r_iR}) = \pm1$ comes from the logarithmic divergence of the Bessel function $K_\frac{D-2}{2}$ when its argument approaches zero. Such situation corresponds to the light particle being at the top of one of the heavy particles.  It is noticeable that at large distances the light particle wave function for $D=2.5$ decreases slower than for $D=3$. This behavior follows from $g(2.5)<g(3)$ in equation~\ref{eq:gD}, which leads to a decrease in the light-heavy dimer binding energy when squeezing the system. In figure~\ref{fig3}~(b), we detail the behavior of the wave function with respect to $r_i/R$ for two fixed values of 
$\cos(\theta_{r_iR})$ (1 and 0.8) and $D$ (2.5 and 3). The results are arbitrarily normalized to $1$ at 
$r_i/R=1$ and $\cos(\theta_{r_iR})=1$. Two properties seen in figure \ref{fig3}~(a) are highlighted in this figure: (i) the peak vanishes fast by decreasing the cosine and (ii) the exponential tail of the wave function is less damped for $D=2.5$ than for $D=3$.

\section{Impurities structure}

In this section, we will study the many-body bound state properties by solving the energy eigenvalue equation for the heavy pair of atoms. By defining a reduced wave function $\chi(R) = R^{(D-1)/2}\phi(R)$ and considering the presence of ($N-2$) light atoms that generate the effective potential between the heavy ones, the energy eigenvalue equation~\eqref{eqheavy} can be rewritten as

\begin{multline}
 \left[-\frac{d^{2}}{dR^{2}}- (N-2) \frac{m_{A}}{\hbar^{2}} \epsilon(R)+\frac{(D-3+2l)(D-1+2l)}{4R^{2}} \right.  \\
 \left.+ \frac{m_{A}}{\hbar^{2}} V_{B}(R)\right] \chi(R) = \frac{m_{A}}{\hbar^{2}} E_{N}^{(D)} \chi(R)
 \, .
 \label{scaleheavy}
\end{multline}

The solutions are determined by the two-body energy $E_2^{(D)}$, the mass ratio $m_B/m_A$, the angular momentum $l$, the effective dimension $D$ and the number of light atoms present in the system. For identical heavy atoms in the same spin state, $l$ will be even for bosons and odd for fermions.

\subsection{ Unitary limit}

First, we investigate the many-body problem under the situation of the Landau fall to the centre~\cite{landau}, which corresponds to the Thomas collapse when the potential has the form $U(r) \approx -\beta/r^{2}\ $, with $\beta > 0$. To achieve that, we take the unitary limit in Eq.~\eqref{assympsmall}, so that the effective potential becomes proportional to $1/R^{2}$ with a strength that depends on the mass ratio $m_B/m_A$ and dimension $D$. In this context, we consider the most excited state, where $E_{N}^{(D)}$ is close to zero, so that the heavy particles eigenvalue equation~\eqref{scaleheavy} becomes 
\begin{multline}
 \left[ -\frac{d^{2}}{dR^{2}}- \frac{m_A}{2\mu_{B,AA}} (N-2) \frac{g(D)}{R^{2}} \right.  \\
+\left. \frac{(D-3+2l)(D-1+2l)}{4R^{2}} \right] \chi(R) = 0\,,\label{landaufall}
\end{multline}
where we have considered null heavy-heavy interaction ($V_B =0$).
Assuming an ansatz $\chi(R) = R^{\delta}$ for the heavy atoms reduced wave function, we can write a second order equation
\begin{multline}
 -\delta(\delta-1)-  \frac{m_A}{2\mu_{B,AA}} (N-2)g(D) \\
 +\frac{(D-3+2l)(D-1+2l)}{4}  = 0\, ,
\end{multline}
whose solutions have the form $ \delta = 1/2 \mp i s$, with
\begin{equation}
 s = \sqrt{\frac{m_A}{2\mu_{B,AA}} g(D) -\frac{(D-2+2l)^{2}}{4}}\, ,
 \label{BOscale}
\end{equation}
where $s$ can be either real or imaginary. For real values of $s$, we have the manifestation of log-periodic oscillations in $\chi(R)$, with the system exhibiting a discrete scale invariance. In the situation where $s$ is purely imaginary, a continuous scale invariance takes place and the wave function has a power-law behaviour instead of the log-periodic one. In such case, the nature of the system belongs to nonrelativistic unparticle physics~\cite{Georgi:2007ek,unnuclear} in the context of cold atoms, which is an interesting and fairly new area of research, and will be explored in future works.

%%%%%%%%%%%%%%%%FIGURE04%%%%%%%%%%%%%%%%%%%
\begin{center}
\begin{figure}[!htb]
\includegraphics[width=7.3cm]{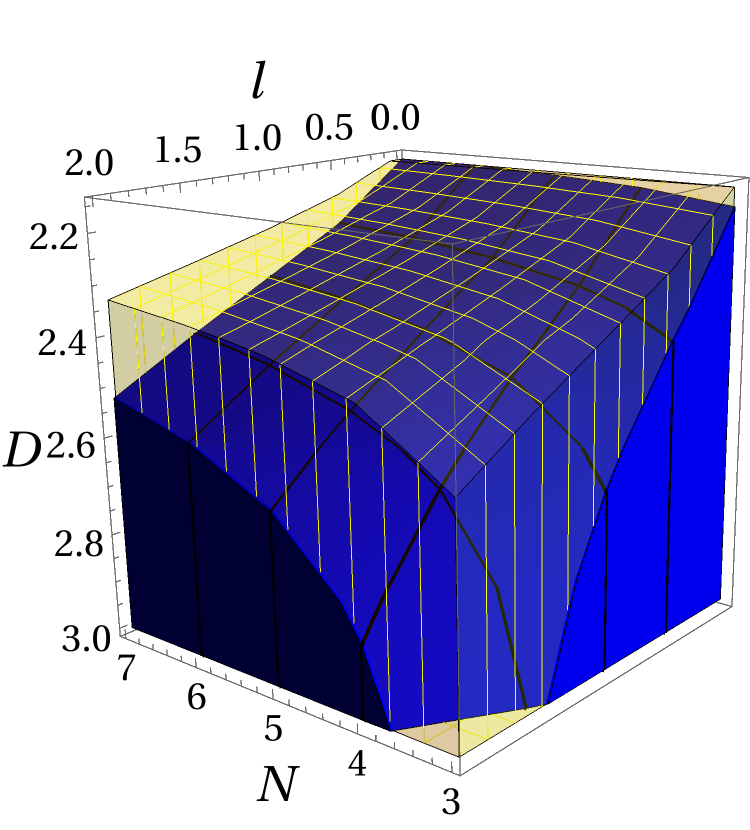}
\caption{Regions for the existence of the Efimov effect in the unitary limit for different dimensions, angular momentum and  total number of atoms. Lower blue and upper yellow regions presents results for $m_B/m_A=6/133$ and $m_B/m_A=0.01$, respectively.}
\label{fig4}
\end{figure}
\end{center}
%%%%%%%%%%%%%%%%%%%%%%%%%%%%%%%%%%%%%%%%%%%%%

To obtain real values of $s$ we have to satisfy the condition
\begin{equation}
  \frac{m_A}{2\mu_{B,AA}} (N-2)g(D) >\frac{(D-2+2l)^{2}}{4}  \ .
  \label{eq:thomascollapse}
\end{equation}
The expression above gives the critical dimensions where the geometrical scaling regime vanishes as a function of $l$ and $N$ for a given mass ratio. In Fig.~\ref{fig4}, it is shown two surfaces, standing for light-heavy mass ratios of 0.01 (upper yellow surface) and 6/133 (lower blue surface). Once the mass ratio decreases for a given number of light atoms, the region for the manifestation of geometrical scaling law enlarges. Besides that, by increasing the number of light bosons, the maximum value of angular momentum that allows the existence of the discrete geometrical scaling parameter increases, while lower values of $D$ becomes accessible.

By replacing the value of $s$ in Eq.~\eqref{landaufall}, we can go forward in the exploration of the many-body system, this time, for finite bound state energies. In this case, we can write a differential equation of the form

\begin{equation}
 \left[ -\frac{d^{2}}{dR^{2}}-  \frac{s^2-1/4}{R^{2}}  \right] \chi(R) = -\kappa_N^{2}\ \chi(R),\label{4.6}
\end{equation}
where $m_A E_N^{(D)}/\hbar^2 = -\kappa_N^{2}$. Assuming bound state solutions and implementing the corresponding boundary condition, one finds that
\begin{equation}
  \chi(R) = \sqrt{R} K_{i s}(\kappa_N R).
\end{equation}
To completely define the solution, we choose the boundary condition where $\chi(R_c)=0$, resulting in discrete values for $\kappa_N$, such that $K_{is}\left(\kappa_N^{(n)} R_c\right)=0$. For shallow bound states, where $\kappa_N^{(n)}R_c \ll1$, the zeros of the Bessel function are given by
\begin{equation}
  \kappa_N^{(n)} R_c= 2e^{-\gamma}\exp{\left(-\frac{n \pi}{s}\right)}[1+ \mathcal{O}(s)],
\end{equation}
where $\gamma$ is the Euler constant. Taking the ratio of consecutive $N$-body energies gives
\begin{equation}
  \frac{E_N^{(n)}}{E_N^{(n+1)}}=\exp{\left(\frac{2\pi}{s}\right)}\ \ \ \ \ n\rightarrow 0,1,2,\cdots\, .
\label{energyscale}
\end{equation}

As we can observe, following a geometrical law like the one found in the Efimov effect, the above equation gives the ratios between successive $N$-body bound energies, even though the system under study is in a generic dimension $D$. In addition, we stress that the scale parameter depends on the number of light bosons, being different from that found in the three-body problem. Therefore, the log-periodic behavior of the wave function and the interwoven limit-cycles~\cite{interwoven} associated with correlations between observables in the $N$-body system built a complex pattern, which survives under squeezing of the system until a critical effective dimension. It is interesting to observe that the log-periodicity of the wave function survives for lower values of 
dimensions when increasing the number of light bosons, such that, in principle, one could arrive to a situation tuned by the trap deformation in which the Efimov effect is vanished, but geometrical scaling laws are still present for more then three particles.

%%%%%%%%%%%%%%%%%FIGURE05%%%%%%%%%%%%%%%%%%%
 \begin{center}
\begin{figure}[!thb]
\subfigure[]{\includegraphics[width=8.55cm]{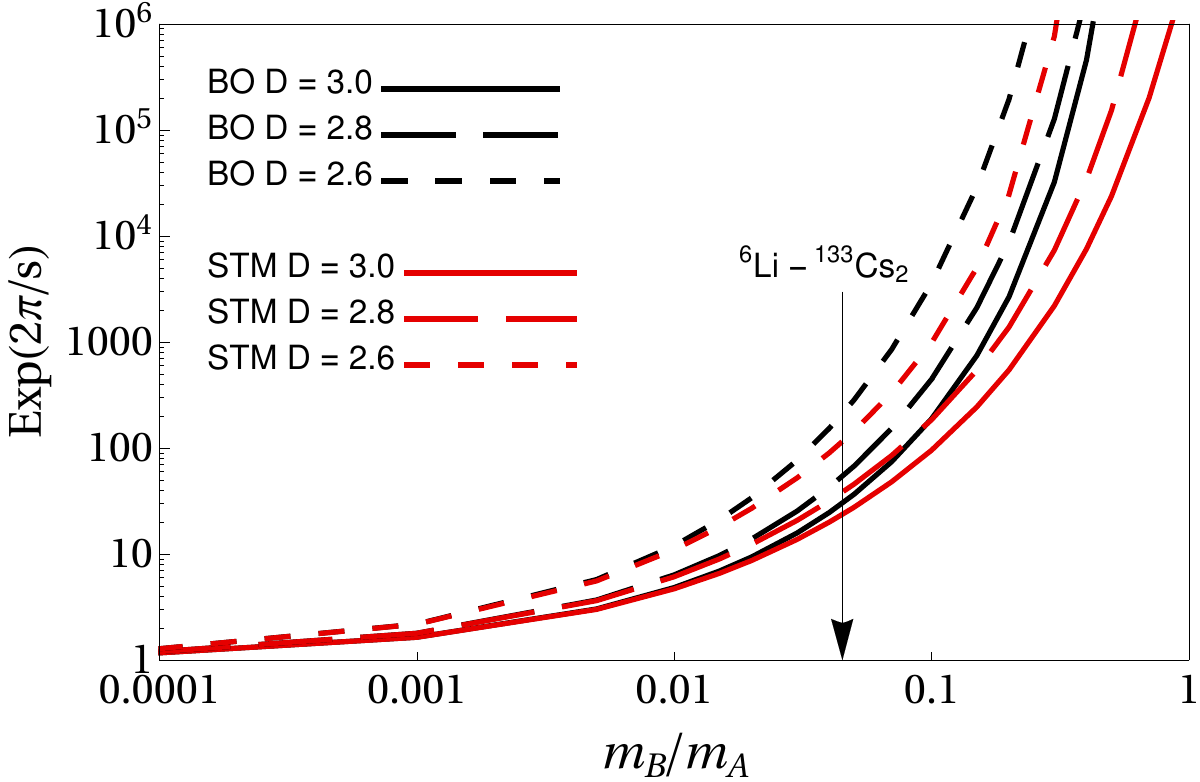}}
\subfigure[]{\includegraphics[width=8.4cm]{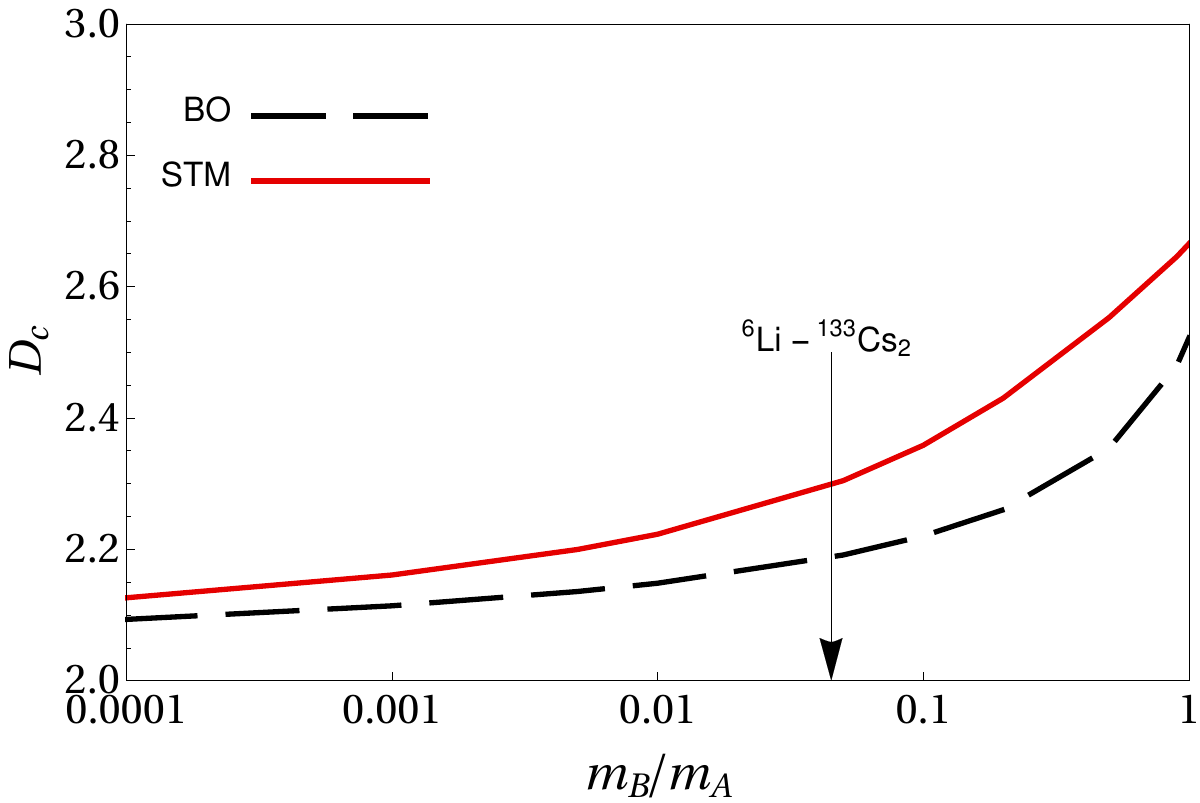}}
\caption{Efimov effect properties for different mass ratios and dimensions for fixed null angular momentum in units of $\hbar=1$.  Figure~(a): Efimov scale parameter in the unitary regime by means of the BO approximation compared to the solutions of Skorniakov-Ter-Martirosyan equations. Figure~(b): Critical dimensions for the manifestation of the Efimov effect as a function of mass ratios.}
\label{fig5}
\end{figure}
\end{center}
%%%%%%%%%%%%%%%%%%%%%%%%%%%%%%%%%%%%%%%%%%%%%

The critical dimension for the manifestation of the Efimov effect has been already investigated by solving the Skorniakov and Ter-Martirosyan equations (STM) at the unitary limit~\cite{rosastm}. Here, we adapt this approach by restricting it to the present situation of only heavy-light atoms resonances to allow comparison with the results from the BO approximation. In the present context, there is only one STM integral equation, which reads
\begin{eqnarray}
 &&\chi_{A}(q) = \tau_{A}\left(E_3- \frac{1}{2\mu_{A,AB}}q^2\right)  \nonumber \\
 &&\times\int d^{D}k \frac{\chi_{A}(k)}{E_3  - (k^2+q^2)/2\mu_{AB} - \mathbf{k}\cdot\mathbf{q}/m_B}\,,   \ \ \ \ \ \ \label{stm}
\end{eqnarray} 
where  $\textbf{q}$ and $\textbf{k}$ are defined such that their origin is the center-of-mass of a given pair and point towards the remaining particle, being  $\tau_A$ the heavy-light transition amplitude. In the limit of large momentum, the integral equation admits solutions for the spectator function in the form  $\chi_A(z)= C_A\ z^{1-D+i s}$, where $C_A$ is the normalization constant.  Substituting this homogeneous function in Eq.~\eqref{stm}, one arrives to the following transcendental equation for $s$
\begin{equation}
\left( \frac{2\mu_{A,AB}}{m_B}\right)
 \left(\frac{\mu_{AB}}{\mu_{A,AB}}\right)^{D/2} =  \mathcal{F}_D  \  I_{(D)}(s)\, ,
\end{equation}
where 
\begin{eqnarray}
\mathcal{F}_D= \frac{1}{\Gamma(D/2-1)\Gamma(2-D/2)}
\end{eqnarray}
and
\begin{small}
\begin{multline}
I_{(D)}(s) = \int^{\infty}_0 dz \, \frac{ 4\, z^{i s}}
{(m_B/m_A) \left(z^2+1\right)+(z-1)^2}  \\
\times\, _2F_1\left(1;\frac{D-1}{2};D-1;-\frac{4 z}{(m_B/m_A) \left(z^2+1\right)+(z-1)^2}\right)\, ,
\end{multline}
\end{small}
with $ _2F_1(a; b; c; d)$ being the ordinary (Gauss) hypergeometric function.

In  Fig.~\ref{fig5}(a), it is compared  the  Efimov geometric ratio calculated by means of the BO and STM equations as a function of the mass ratio for different dimensions.  As expected, the results of the BO approximation improves as the heavy-light mass ratio increases. Besides that, we find that the BO approximation overestimate the geometric ratio, turning the effective potential in equation~\ref{4.6} less attractive, while such behavior is reversed when approaching the critical dimension, as observed in Fig.~\ref{fig5}(b). These differences occur mainly because of the assumptions made in the two methods, i.e, in the BO, the heavy atoms degrees of freedom are frozen, while in the STM approach all bosons are free to move. Finally, considering a three-body system formed by $^6$Li-$^{133}$Cs$_2$, we found that the critical dimension shown in Fig.~\ref{fig5}(b) is underestimated on the order of $5\%$.

\subsection{Finite range corrections}
\label{ImpBoundStates}

Experimentally, the access to regimes where the system exhibits discrete scaling symmetry is made by controlling the scattering length via Feshbach resonances, so it is interesting to write the two-body energy in terms of the respective scattering length in a generic dimension $D$. Besides that, within the real possibilities of experimental cold-atom setups, the unitary limit is elusive, such that one has to consider finite scattering lengths in order to obtain the properties of the $N$-body system close to the Feshbach resonances. Taking this into account, our study is now devoted to quantify corrections out of the unitary limit while varying the effective dimension in which the system is embedded. In order to accomplish this, we consider the heavy-light pair in our system to be two non-relativistic particles with a rotationally-invariant zero-range interaction. In this context, we follow closely Ref.~\cite{2bodies}. 

In the zero-range model, the radial wave function for $r>0$ is
\begin{eqnarray}
R(r) &=& \sqrt{\frac{\pi}{2p}} r^{\frac{2-D}{2}} \left[\cot\delta_{(D)}(p)\ J_{\frac{D}{2}-1}(p\, r) \right. \nonumber \\ 
&-& \left.
Y_{\frac{D}{2}-1}(p\, r) \right]\, ,
\end{eqnarray}
where $r$ is the distance between the pair of particles and $p^2/2\mu_{AB}$ is the relative energy. The phase-shift is written in terms of the scattering length by assuming that the wave function vanishes at the relative distance $a_D$, which gives
\begin{eqnarray}
\cot\delta_{D}(p) =Y_{\frac{D}{2}-1}(p\, a_{D})/J_{\frac{D}{2}-1}(p\, a_{D}).
\end{eqnarray}
For small energies, we can write
\begin{eqnarray}
\cot\delta_{D}(p) &=&\cot \left(\frac{D }{2}\pi\right)-\frac{2^{D-2}  }{\pi }\Gamma \left(\frac{D-2}{2}\right)  \nonumber \\
&\times&\Gamma\left(\frac{D}{2}\right)\,(a_{D}\ p)^{2-D}.\label{eq:cotdelta}
\end{eqnarray}
 The S-matrix in $D$-dimensions is written as
\begin{eqnarray}
{S} = {1}+ \left(\frac{i \ p} {2\pi}\right)^{\frac{D-1}{2}}{f}_{D}(p),\ \ \ \
\end{eqnarray}
where the scattering amplitude is given by
\begin{equation}
{f}_{D}(p) = \frac{1}{p^{D-2}}\frac{1}{\cot{\delta_{D}(p)}-i}\, .
\end{equation}
The two-body binding energies are located at the poles of the S-matrix in the real axis. In the lowest order of the effective range expansion, equation~\eqref{eq:cotdelta} gives 
\begin{equation}
p=\frac{1}{a_{D}} \left(\frac{2^{D-2}}{\pi}\frac{\Gamma \left(D/2-1\right) \Gamma \left(D/2\right) }{\cot \left(D\pi/2\right)-i}\right)^{\frac{1}{D-2}}\,.\label{eq:e2ad}
\end{equation}
Being  $p=i \kappa_2$ with $\kappa_2>0$, we can write $E^ {(D)}_2=-\hbar^2 \kappa_2^2/2\mu_{AB}$,
which reproduces the results in reference~\cite{molmerd}.

%%%%%%%%%%%%%%%%FIGURE06%%%%%%%%%%%%%%%%%%%
\begin{center}
\begin{figure}[!thb]
\includegraphics[width=8.5cm]{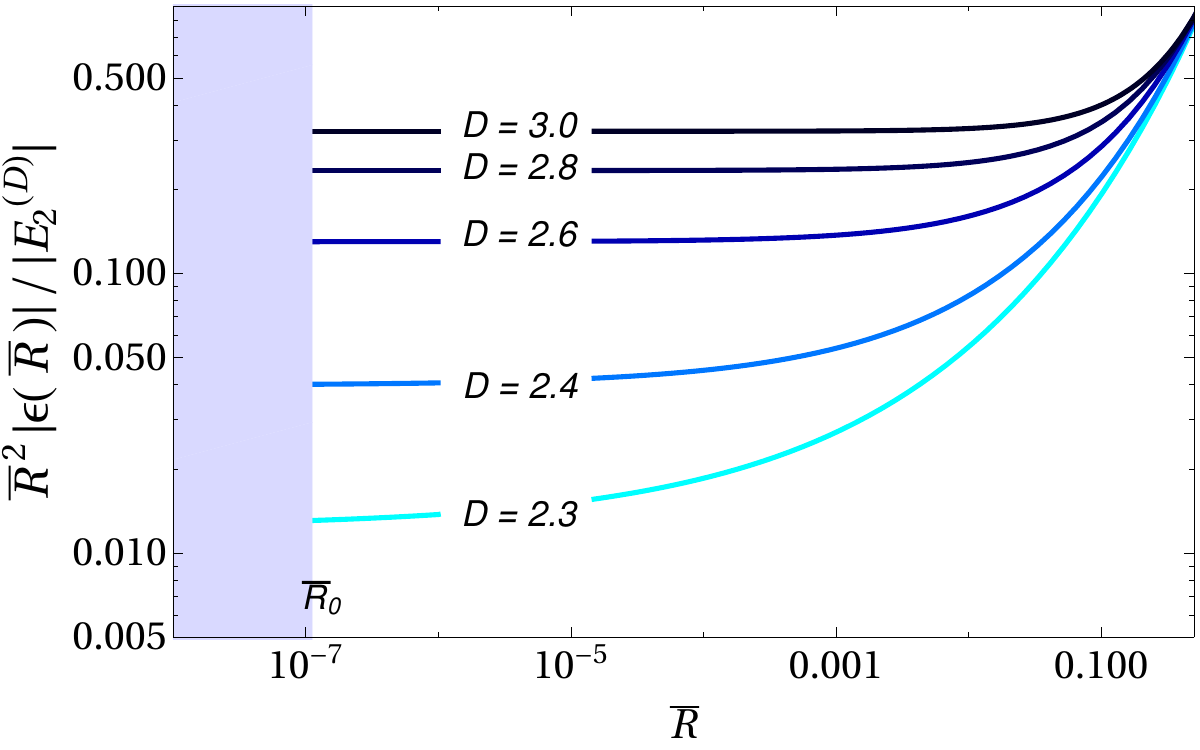}
\caption{Effective potential for different dimensions $D$  as a function of $\overline R=R/L$ $(L=[ \hbar^{2}/{2\mu_{B,AA}|E^{(D)}_{2}|}]^\frac12)$. The band corresponds to the region where the  short-range potential is repulsive to avoid the Thomas collapse. }
\label{fig6}
\end{figure}
\end{center}
%%%%%%%%%%%%%%%%%%%%%%%%%%%%%%%%%%%%%%%%%%%%%

In order to obtain the bound states of the two heavy boson impurities in the cloud of $(N-2)$ light atoms, the effective potential obtained from equation~\eqref{fullbopot} has to be regularized at short distances, which is done by associating a distance $R_0$, named van der Walls length, with the repulsive region of real potentials. This is necessary 
to avoid the Thomas collapse when the condition given by Eq.~\eqref{eq:thomascollapse} is satisfied. As an example, in cold-atomic systems, $R_0$ is of the order of the Van der Waals radius, which determines to some extend the position of the three-atom recombination resonance, as in the case of three identical atoms~\cite{JohansenNatPhys2017}. 
Within this approach, the effective potential becomes
\begin{equation}
\epsilon( R) \rightarrow\textrm{sign}\left( R_0- R\right)\epsilon( R)\,, \end{equation}
obtained by an appropriate choice of $V_B(R)$ in  the eigenvalue equation~\eqref{scaleheavy}.

In Fig.~\ref{fig6}, we illustrate the effective potential focusing on the $1/R^2$ behavior for different values of effective dimension $D$. As we can observe, the region where the effective potential exhibits the $1/R^2$ behavior diminishes as the effective dimension is decreased, so that, in principle, as the trap geometry asymmetry increases, the ratio $R_0/a_D$ has to be tuned to lower values in order to accesses the geometrical scaling regime. 

%%%%%%%%%%%%%%%%%FIGURE07%%%%%%%%%%%%%%%%%%%
\begin{center}
\begin{figure}[!thb]
\subfigure[]{\includegraphics[width=8.5cm]{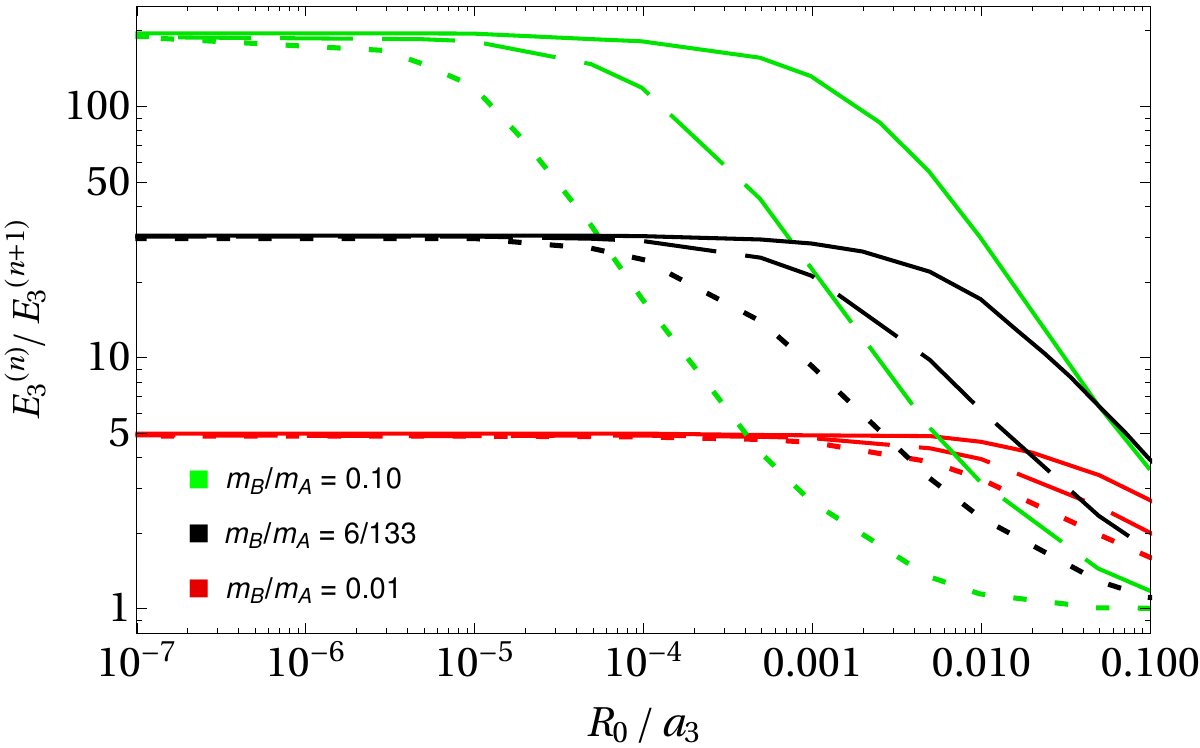}}
%\vspace{-0.3cm}
\subfigure[]{\includegraphics[width=8.5cm]{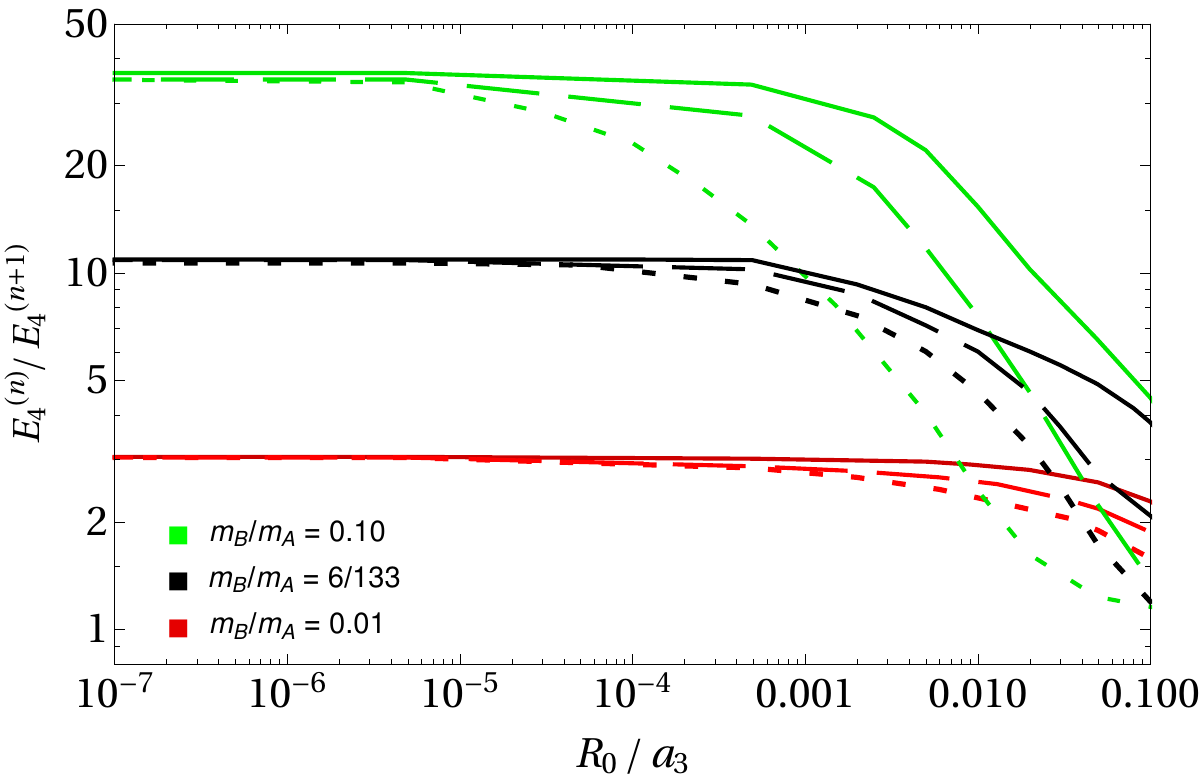}}
\caption{ Ratio between the energies of two successive states, $E_N^{(n)}/E_N^{(n+1)}$, as a function of $R_0/a_3$ for dimension $D=3$ and angular momentum $l=0$. We consider two heavy bosonic impurities interacting with one ($^6$Li$-\,^{133}$Cs$_2$) and two ($^6$Li$_{2}-\,^{133}$Cs$_2$) light bosons, represented respectively in figures (a) and (b)}. The curves from top to bottom are obtained for mass ratios of  $m_B/m_A=0.1$, $6/133$ and $0.01$ and the solid lines corresponds to the excitation quantum number $n=0$, the long-dashed $n=1$ and the short-dashed $n=2$. 
\label{fig7}
\end{figure}
\end{center}
%%%%%%%%%%%%%%%%%%%%%%%%%%%%%%%%%%%%%%%%%%%%%

 %%%%%%%%%%%%%%%%%FIGURE08%%%%%%%%%%%%%%%%%%%
\begin{center}
\begin{figure}[!thb]
\subfigure[]{\includegraphics[width=8.5cm]{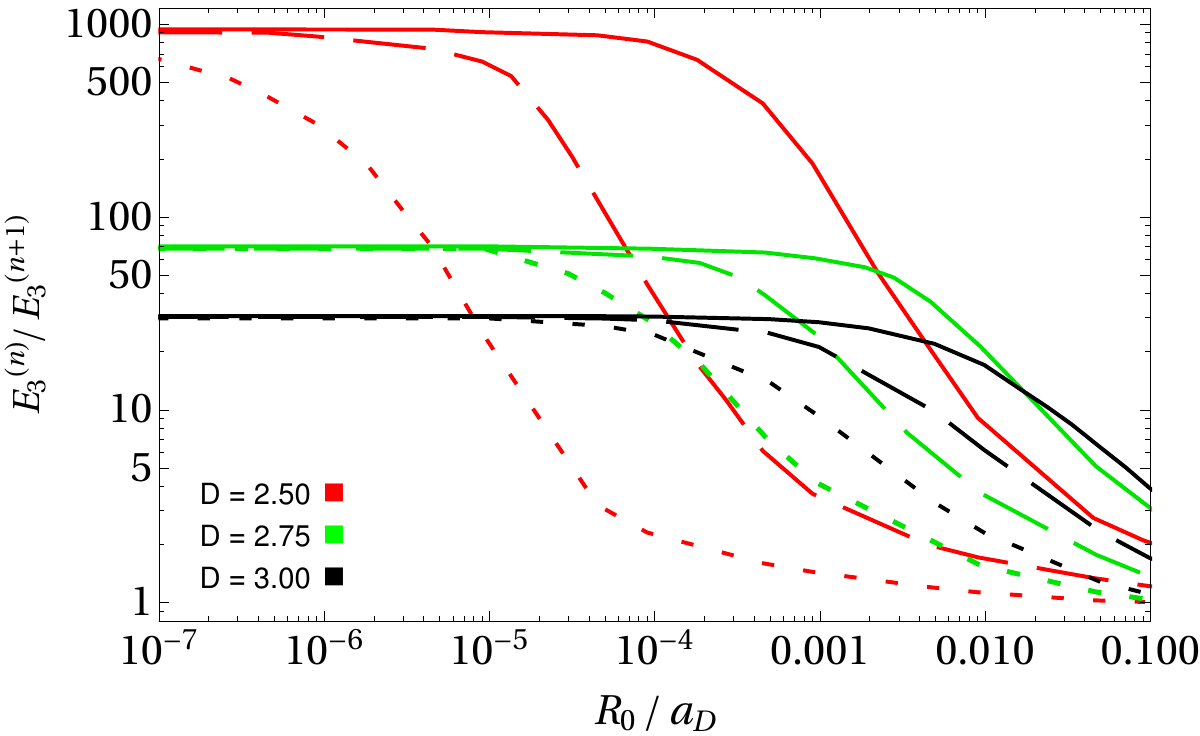}}
\subfigure[]{\includegraphics[width=8.5cm]{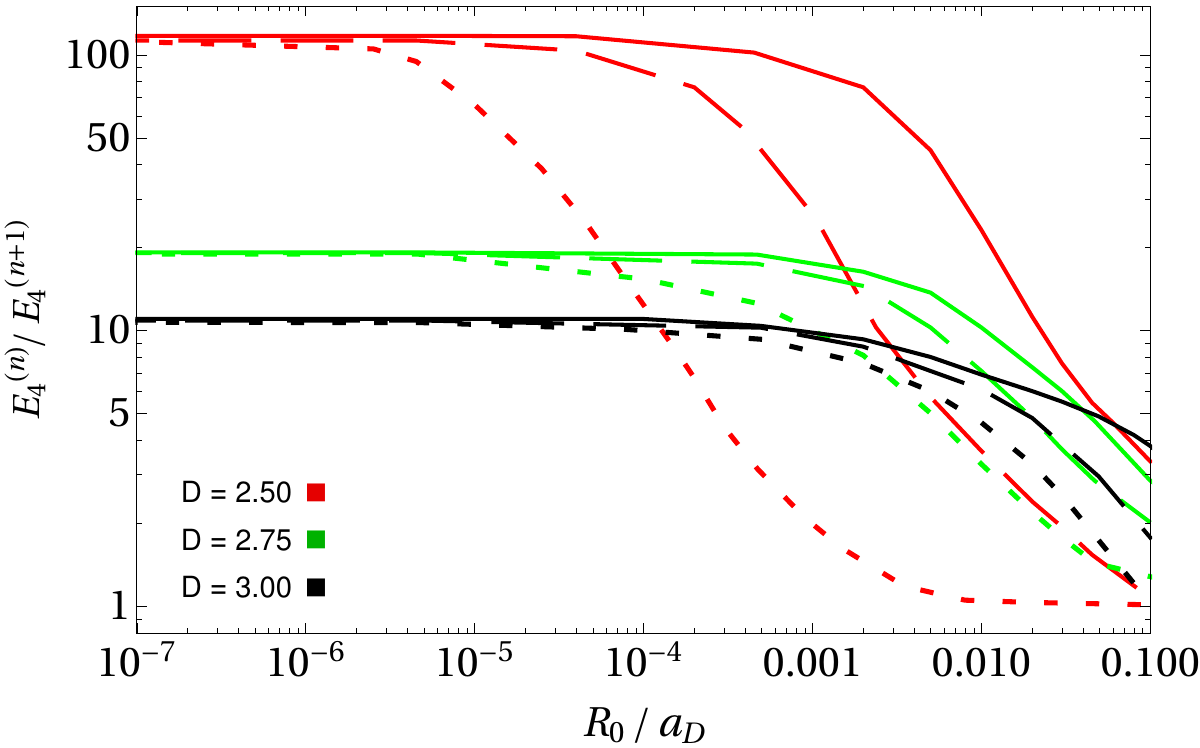}}
\caption{
Ratio between the energies of two successive states, $E_N^{(n)}/E_N^{(n+1)}$, as a function of $R_0/a_D$  for fixed angular momentum $l=0$. We consider two atoms of $^{133}$Cs interacting with one and two $^6$Li ones, represented respectively in figures (a) and (b). The solid lines corresponds to the excitation quantum number $n=0$, the long-dashed $n=1$ and the short-dashed $n=2$. The curves from top to bottom are obtained for dimensions $2.5$, $2.75$ and $3$.}
\label{fig8}
\end{figure}
\end{center}
%%%%%%%%%%%%%%%%%%%%%%%%%%%%%%%%%%%%%%%%%%%%%

In figure~\ref{fig7}, we present the ratio of two successive bound state energies as a function of $R_{0}/a_{3}$ for (a) three- and (b) four-body systems embedded in three dimensions for some mass ratios. As shown in the diagrams, when $a_{3}\gg R_0$, the bound states present a geometric ratio of the form $\exp(2\pi/s)$ that can be represented by a unique limit-cycle. Besides that, we can see that the magnitude of the scattering length necessary to obtain discrete scale symmetry is larger for the systems that present greater light-heavy mass ratios. This can be explained by means of the strength of the effective potential between the heavy atoms generated by the presence of the light ones, which is proportional to the inverse of the reduced mass $\mu_{B,AA}$ (please see Eq.~\eqref{assympsmall}. Finally, as one sees by comparing figures \ref{fig7}(a) and \ref{fig7}(b), the ratio between the energy of successive states depends on the number of light-bosons. This property was also found in the four-boson system with zero-range interactions~\cite{hadizadeh2011}, and in the present model it comes from the increase of the strength of the inverse square potential in the BO approach by increasing the number of light atoms.

In Fig.~\ref{fig8}, the ratio between energies of two successive states as a function of $R_0/a_D$ are presented for  (a) $^6$Li-$^{133}$Cs$_2$ and (b) $^6$Li$_2$-$^{133}$Cs$_2$ molecules. Considering the effective dimensions 3, 2.75 and 2.5, with zero angular momentum, we have analysed up to three consecutive states, with $n=0$, 1 and 2. As we can observe in both figures, as the effective dimension decreases, the geometric ratio between successive energies increases until the critical dimension is reached. Comparing Figs.~(a) and (b), one can observe that independent of the dimension, the discrete scale regime of the
$^6$Li$_2$-$^{133}$Cs$_2$ molecule is more resilient against the increasing of $R_0/a_D$ when compared to the $^6$Li-$^{133}$Cs$_2$ one. This can be understood since the $1/R^2$ potential has twice the strength in the former case compared to latter one. Another aspect observed in both figures that is also independent on the effective dimension, is the behaviour of the $n+1$ energy  state that changes in magnitude slowly with respect to the $n$ one when the ratio $R_0/a_D$ is increased. The reason for that is in the form of the effective potential, which is deep and goes with the behaviour $1/R^2$ at short distances and is exponentially damped at large separations of the two heavy atoms. The increase of $R_0/a_D$ raises the kinetic energy, while pushing the states to the shallow region of the potential. The excited state, although gaining kinetic energy, does not loose considerable potential energy, while the lower state, which initially is more deep and localized in the $1/R^2$ region, gain less kinetic energy, while looses more potential energy when it is moved to the region of the shallow potential.  We should observe that the departure from the geometrical scaling regime becomes more dramatic when $D$ is decreased.  The reason for that is traced back to Fig.~\ref{fig6}, which shows that the region where the potential $1/R^2$ ($\overline R\simeq R/a_D$) acts is narrowed by decreasing $D$.   

\begin{figure}[!thb]
\subfigure[]{\includegraphics[width=8.5cm]{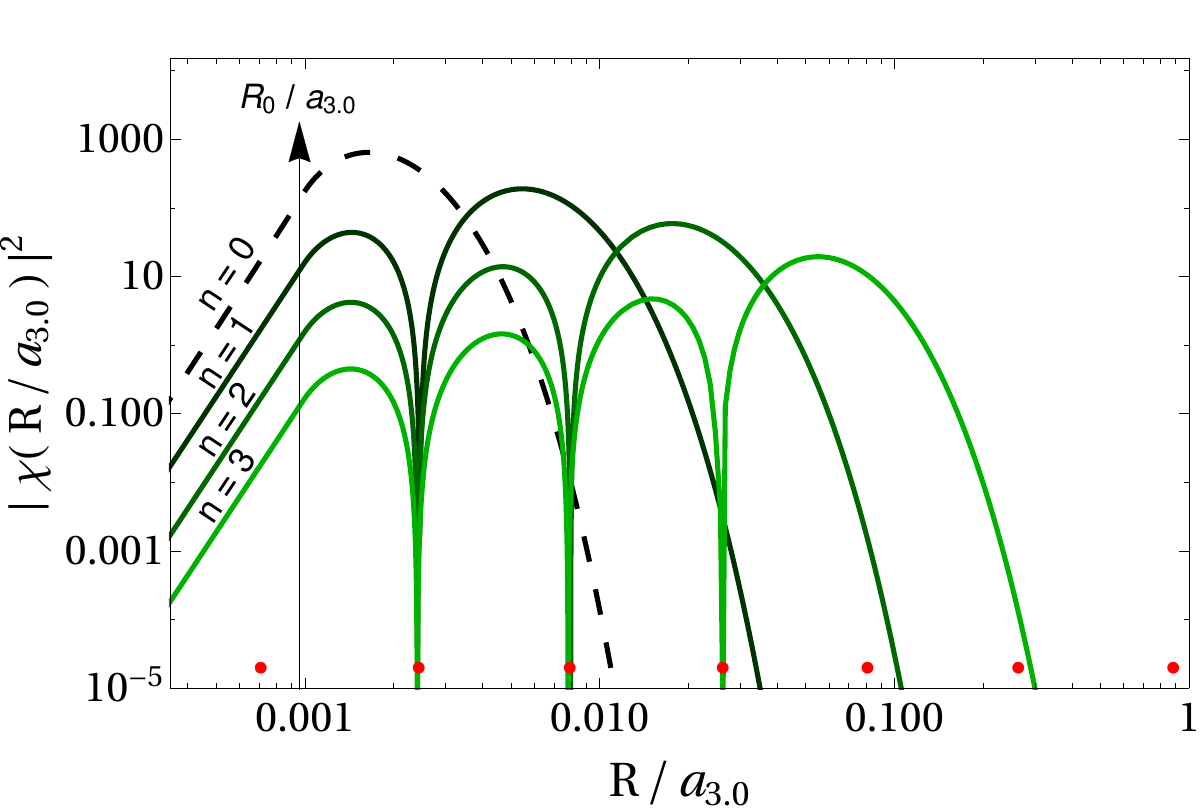}}
\subfigure[ ]{\includegraphics[width=8.5cm]{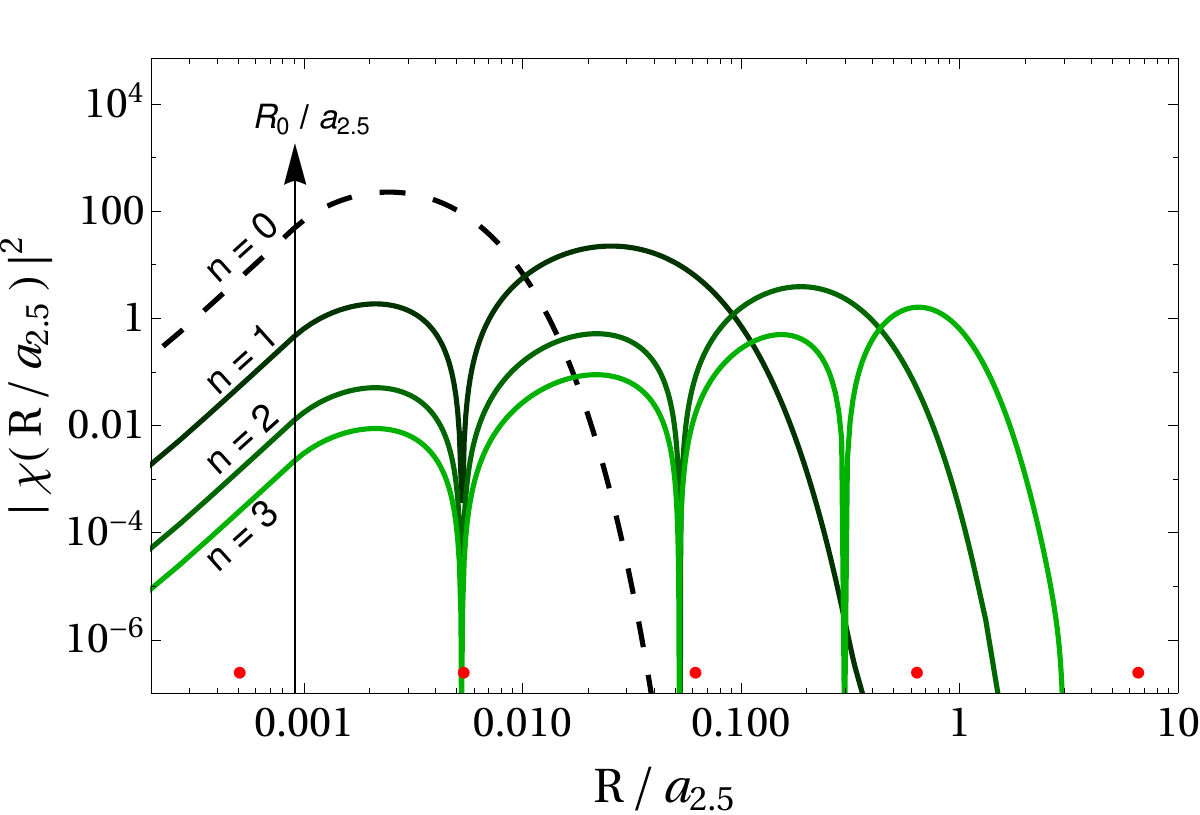}}
\caption{We computed the squared modulus of the bound state wave function of the heavy-pair, $|\chi(R/a_{D})|^{2}$, in the lowest angular momentum state 
considering the molecule $^{133}$Cs$_2$-$^6$Li$_2$ embedded in two different dimensions, $D=3$ and $2.5$, represented in figures (a) and (b), respectively.  The cutoff radius in units of the scattering length is fixed in  $R_0/a_3=9\times 10^{-4}$. The ground state $n=0$ is the dashed line, the solid lines from top to bottom are $n=1$, 2 and 3. The dots (red) represent the nodes of the highest excited state when the system is at the unitary limit.}
\label{fig9}
\end{figure}

In what follows, we stress the relation of the scattering length with the effective dimension studying the wave function of the heavy-heavy pair. In Fig.~\ref{fig9}, for different effective dimensions (a) $D=3$ and (b) 2.5, we illustrate the square modulus of the wave function of the pair $^{133}$Cs-$^{133}$Cs in the molecule $^6$Li$_2$-$^{133}$Cs$_2$ for $n=0$, 1, 2 and 3. The dots (red) represent the nodes of the highest excited state when the system is at the unitary limit. These nodes can be used as a reference on how close to the log-periodicity behavior the wave function is. By comparing the figures, we can observe that changing the effective dimension from $D=3$ to $D=2.5$, for a fixed ratio $R_0/a_D$, makes the deviation from the logarithmic periodicity regime greater, which highlights the relation between the magnitude of the ratio $R_0/a_D$ and how close the many-body energies are from the geometrical scaling regime.

%%%%%%%%%%%%%%%%%FIGURE09%%%%%%%%%%%%%%%%%
 
%%%%%%%%%%%%%%%%FIGURE10%%%%%%%%%%%%%%%%%%%
\begin{center}
\begin{figure}[!thb]
\includegraphics[width=8.5cm]{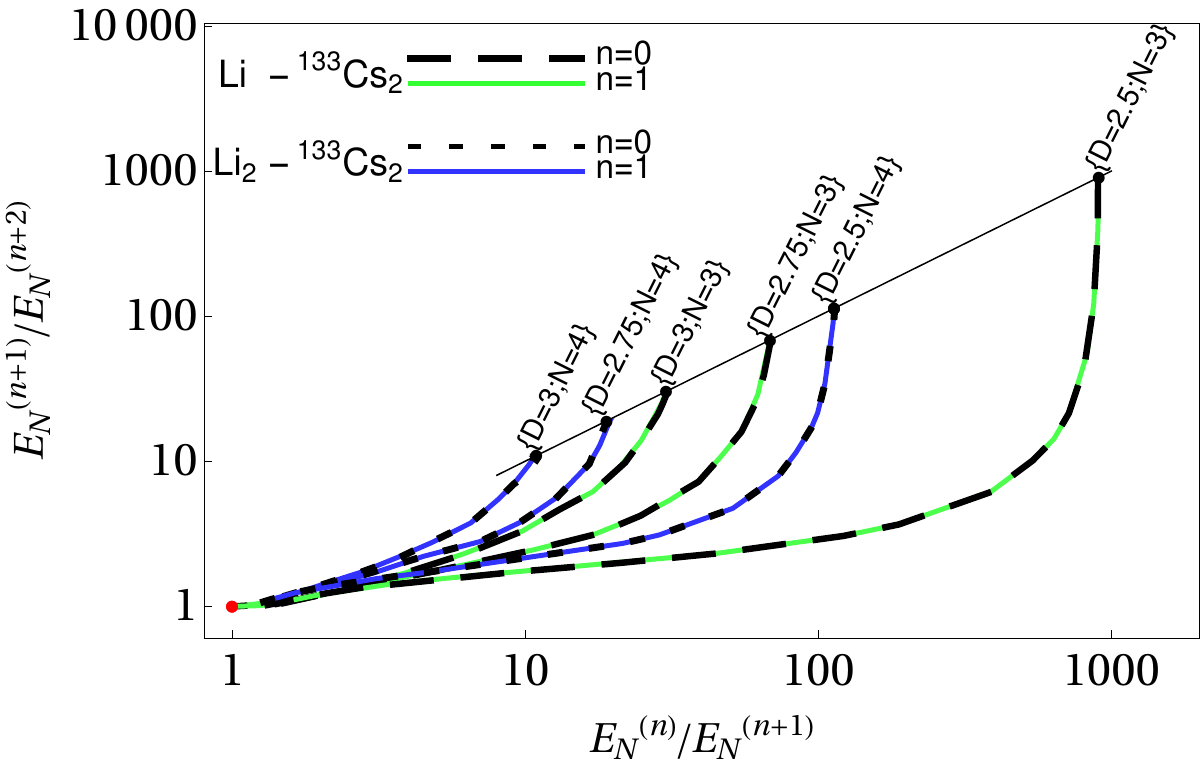}
\caption{Results for the scaling function $E^{(n+1)}_N/E^{(n+2)}_N$ vs.
$E^{(n)}_N/E^{(n+1)}_N$ considering the molecules $^6$Li-$^{133}$Cs$_2$ and 
$^6$Li$_2$-$^{133}$Cs$_2$ for $n=0$ and $n=1$. The three- and four-body system is squeezed in different effective dimensions. The straight line is the unitary limit. The red dot is the limit where the states reaches the continuum. }
\label{fig10}
\end{figure}
\end{center}

To finalize the analysis of the many-body bound states, in figure~\ref{fig10} we show the scaling functions $E^{(n+1)}_N/E^{(n+2)}_N$ vs.
$E^{(n)}_N/E^{(n+1)}_N$ for the $^6$Li-$^{133}$Cs$_2$ and $^6$Li$_2$-$^{133}$Cs$_2$ molecules embedded in different dimensions, with the bound-state energies computed by changing the ratio $R_0/a_D$. Two scaling functions for $n=0$ and $n=1$ are computed, the results feature the limit-cycles for three and four particles. The black solid line denotes the geometric scaling ratio at the unitary limit and the red dot represents the limit where the energies of the three- and four-body systems become equal to the two-body energy and twice it, respectively, reaching the continuum at this point. The limit-cycles are characteristic of the dimension and also the number of light bosons.  It is worthwhile to mention that the detailed form of the correlation
between $E^{(n+1)}_N/E^{(n+2)}_N$ and
$E^{(n)}_N/E^{(n+1)}_N$ depends on the effective potential, $\epsilon(R)$, 
which varies from the short distance behavior, $1/R^2$~\eqref{assympsmall}, to the long distance one with an exponential tail proportional to the two-body energy~\eqref{assymplarge}.

\section{SUMMARY}
\label{conclusion}

In this work, we have explored the problem of two heavy bosonic impurities immersed in a system of light bosons by means of the Born-Oppenheimer approximation. The many-body system is embedded in $D$ dimensions in order to mimic squeezed traps. Correlations between the many-body bound state energies and wave function for different dimensions, mass ratio and number of light bosons are discussed. The study have focused in the regime of discrete scale symmetry up to the point of transition to the continuum one, considering both zero and finite heavy-light two-body bound energies. 

We have placed our study in the context of the $^6$Li$_{N-2}\,-^{133}$Cs$_2$ molecule, where the  $^6$Li$-^6$Li interaction is weak, which would be the case when the scattering length vanishes. In the particular case of $N=3$ by means of the Born-Oppenheimer approximation in different dimensions, we computed for the lowest angular momentum state at the unitary limit the critical dimension where the Efimov effect vanishes ($D_c\simeq2.188$), which represents an overestimation of 5\% over the Skorniakov and Ter-Martirosian result ($D_c\simeq2.298$). 
For $N=3$ and $N=4$, we also study  the $^{133}$Cs-$^{133}$Cs wave function up to three excited states for some effective dimensions and different ratios between the short-range cutoff ($R_0$) and the heavy-light scattering length ($a_D$). Besides that, we have found that the tune of $R_0/a_D$, necessary to place the system into the discrete scale regime, is strongly correlated to the effective dimension in which the system is embedded. 

Furthermore, we have found that the effective dimension where the discrete scale symmetry breaks to the continuum one depends on several system variables, such as, the angular momentum, number of light bosons and mass imbalance between heavy-light atoms. As a consequence of this, we can find situations in squeezed traps where systems with three bosons are out of the discrete scaling symmetry regime, while four or more bosons are still in. Furthermore, we have found that an increase in the angular momentum of the heavy-heavy pair, which would tends to destroy the discrete scale symmetry regime, can be balanced by increasing the number of light bosons in order to preserve discrete scaling properties. 

Below the critical dimension where the discrete scaling symmetry ceases to exist and the continuum one takes place, a
phenomenon closely related to the unparticle physics~\cite{Georgi:2007ek} can be explored. Already studied in the context of nuclear physics~\cite{unnuclear},  the key to observe the unparticle physics in cold atomic traps is the presence of a power-law behavior with a nontrivial exponent related to the few-atom recombination rate. In reference~\cite{prlunparticle} the possible manifestation of anatomic physics in cold atom physics
was already pointed out, which gives interesting features to take into
account when planning and developing future experiments with the aim
of observing the transition between the discrete and continuum scale
symmetry regimes. 

In summary, our study corroborates with the existence of geometrical scaling laws for $N$-body systems with different log-periodicity properties for each system configuration. Such properties were also found in Ref.~\cite{hadizadeh2011} for four identical atoms with non-zero interactions in momentum space. Hence, for future works, we plan to introduce a non-zero interaction between the light particles into our system in coordinate space, which makes the problem much more difficult to solve in mathematical terms and could bring some corrections to the quantities that we have found in the present paper.  At least, in principle, the three-body threshold that appears when the light particles interact between themselves should not affect the deeper four-boson bound states in the discrete scale regime, once it is located below this threshold. Indeed, this has been verified for four identical bosons in reference~\cite{hadizadeh2011}.  Besides that, the physics of few-atom recombination rates below the critical dimension, where non-relativistic unatomic phenomena is expected to arise, looks very interesting and we also aim to investigate it in future works.

\section*{ACKNOWLEDGMENTS}

This work was partially supported by Funda\c{c}\~{a}o de Amparo \`{a} Pesquisa do Estado de S\~{a}o Paulo (FAPESP) [Grants No. 2017/05660-0 (T.F.) and No. 2020/00560-0 (D.S.R.)], Conselho Nacional de Desenvolvimento Cient\'{i}fico e Tecnol\'{o}gico (CNPq) [Grants No. 308486/2015-3 (T.F.)], and Funda\c{c}\~{a}o de Amparo \`{a} Pesquisa do Estado de Minas Gerais (FAPEMIG) [Grants No. 11608].


\begin{thebibliography}{40}

\bibitem{efimov} V. Efimov, Phys. Lett. B \textbf{33}, 563 (1970).

\bibitem{kraemer} T. Kraemer, M. Mark, P. Waldburger, J. G. Danzl, C. Chin, B. Engeser, A. D. Lange, K. Pilch, A. Jaakkola, H.-C. N\"{a}gerl, and R. Grimm,  Nature (London) \textbf{440}, 315 (2006).

\bibitem{homoexp1} B. Huang, L. A. Sidorenkov, R. Grimm, and J. M. Hutson, Phys. Rev. Lett. \textbf{112}, 190401 (2014).

\bibitem{homoexp2} J. R. Williams, E. L. Hazlett, J. H. Huckans, R. W. Stites, Y. Zhang, and K. M. O'Hara,
Phys. Rev. Lett. \textbf{103}, 130404 (2009).


\bibitem{heteexp1} R. Pires, J. Ulmanis, S. H\"{a}fner, M. Repp, A. Arias, E. D. Kuhnle, and M. Weidem\"{u}ller, Phys. Rev. Lett. \textbf{112}, 250404 (2014).

\bibitem{heteexp2} S.-K. Tung, K. Jim\'{e}nez-Garc\'{i}a, J. Johansen, C. V. Parker, and C. Chin, Phys. Rev. Lett. \textbf{113}, 240402 (2014).

\bibitem{heteexp3} R. S. Bloom, M.-G. Hu, T. D. Cumby, and D. S. Jin, Phys. Rev. Lett. \textbf{111}, 105301 (2013).

\bibitem{efimovatoms} E. Braaten and H.-W. Hammer, Annals Phys. \textbf{322}, 120
(2007).

\bibitem{efimovpolarons} M. Sun, H. Zhai, and X. Cui, Phys. Rev. Lett. \textbf{119}, 013401
(2017).

\bibitem{efimovdipolar} S. Moroz, J. P. D’Incao, and D. S. Petrov, Phys. Rev. Lett. \textbf{115}, 180406 (2015).

\bibitem{efimovphotons} M. J. Gullans, S. Diehl, S. T. Rittenhouse, B. P. Ruzic, J. P. D’Incao, P. Julienne, A. V. Gorshkov, and J. M. Taylor, Phys. Rev. Lett. \textbf{119}, 233601 (2017).
 
\bibitem{nielsen} E. Nielsen, D. V. Fedorov, A. S. Jensen, and E. Garrido, Phys. Rep. \textbf{347}, 373 (2001). 
 
\bibitem{Braaten:2004rn} E.~Braaten and H.-W.~Hammer, Phys. Rep. \textbf{428}, 259 (2006).


\bibitem{Frederico:2012xh}
T.~Frederico, A.~Delfino, L.~Tomio, and M.~T.~Yamashita, Prog. Part. Nucl. Phys. \textbf{67}, 939 (2012).


\bibitem{Naidon:2016dpf} P.~Naidon and S.~Endo, Rep. Prog. Phys. \textbf{80}, 056001 (2017).


\bibitem{Greene:2017cik} C.~H.~Greene, P.~Giannakeas, and J.~Perez-Rios, Rev. Mod. Phys. \textbf{89},  035006 (2017).
 
\bibitem{thomas} L. H. Thomas, Phys. Rev. \textbf{47}, 903 (1935).


\bibitem{rosastm} D. S. Rosa, T.~Frederico, G. Krein, and M. T. Yamashita, Phys. Rev. A \textbf{97}, 050701(R) (2018). \textit{Erratum}, Phys. Rev. A \textbf{104}, 029901 (2021).

\bibitem{mohapatra} A. Mohapatra and E. Braaten, Phys. Rev. A \textbf{98}, 013633 (2018).

\bibitem{cristensen} E. R. Christensen, A. S. Jensen, and E. Garrido, Few-Body Syst. \textbf{59}, 136 (2018).

\bibitem{garridoprr} E. Garrido and A. S. Jensen, Phys. Rev. Research \textbf{2}, 033261 (2020).


\bibitem{BEC2D} D. S. Petrov, M. Holzmann, and G. V. Shlyapnikov, Phys. Rev. Lett. \textbf{84}, 2551 (2000).

\bibitem{BEC1D}  M. Greiner,  I. Bloch, O. Mandel, T. W. H\"{a}nsch, and T. Esslinger, Appl. Phys. B 
\textbf{73}, 769 (2001).

\bibitem{kroger} H. Kr\"{o}ger and R. Perne, Phys. Rev. C \textbf{22}, 21 (1980).

\bibitem{adhikari} S. K. Adhikari and A. C. Fonseca, Phys. Rev. D \textbf{24}, 416 (1981).

\bibitem{naus}  H. W. L. Naus and J. A. Tjon, Few-Body Syst. \textbf{2}, 121 (1987).

\bibitem{yamashita81} M. T. Yamashita, D. V. Fedorov, and A. S. Jensen, Phys. Rev. A \textbf{81}, 063607 (2010).

\bibitem{yamashita75} M. T. Yamashita, L. Tomio, A. Delfino, and T. Frederico,  Europhys. Lett. \textbf{75}, 555 (2006).

\bibitem{von} J. von Stecher,  Phys. Rev. Lett. \textbf{107}, 200402 (2011).

\bibitem{yan} Y. Yan and D. Blume, Phys. Rev. A \textbf{92}, 033626 (2015).

\bibitem{hadizadeh2011} M. R. Hadizadeh, M. T. Yamashita, L. Tomio, A. Delfino, and T. Frederico, Phys. Rev. Lett. \textbf{107}, 135304 (2011).


\bibitem{NaidonFBS2018} P. Naidon, Few-Body Syst. \textbf{59}, 64 (2018).

\bibitem{hadizadeh2013} M. R. Hadizadeh, M. T. Yamashita, L. Tomio, A. Delfino, and T. Frederico, Phys. Rev. A \textbf{87}, 013620 (2013).


\bibitem{interwoven} W.~De Paula, A.~Delfino, T.~Frederico, and L.~Tomio, J. Phys. B: At., Mol. Opt. Phys. \textbf{53}, 205301 (2020).



\bibitem{rosaBO} D. S. Rosa, T. Frederico, G. Krein, and M. T. Yamashita, J. Phys. B: At., Mol. Opt. Phys. \textbf{52}, 025101 (2018).

\bibitem{landau} L. D. Landau, and E. M. Lifshitz,  \textit{Quantum Mechanics} (Pergamon Press, London, 1977).



\bibitem{Georgi:2007ek} H.~Georgi, Phys. Rev. Lett. \textbf{98}, 221601 (2007).

\bibitem{unnuclear} H.-W.~Hammer and D.~T.~Son, Proc. Natl. Acad. Sci. U.S.A. \textbf{118}, e2108716118 (2021).

\bibitem{2bodies} H.-W. Hammer and D. Lee, Phys. Lett. B \textbf{681}, 500 (2009).


\bibitem{molmerd} M. Valiente, N. T. Zinner, and K. M\text{ø}lmer, Phys. Rev. A \textbf{86}, 043616 (2012).

\bibitem{JohansenNatPhys2017} J. Johansen, B. J. Desalvo, K. Patel, and C. Chin, Nature Physics \textbf{13}, 731 (2017).


\bibitem{prlunparticle} E. Braaten and H.-W. Hammer, Phys. Rev. Lett. \textbf{128}, 032002 (2022).


\end{thebibliography}
\end{document}